\documentclass[sigconf]{acmart}

\acmConference[MSR 2024]{21st International Conference on Mining Software Repositories}{April 2024}{Lisbon, Portugal}

\usepackage{geometry}
\usepackage{textcomp}
\usepackage{pgfplots}
\pgfplotsset{width=9cm,compat=1.9}

\usepackage{lipsum}
\usepackage{rotating}
\usepackage{lscape}
\usepackage{listings}
\usepackage{algorithmic}
\usepackage{graphicx}
\usepackage{textcomp}
\usepackage{xcolor}
\usepackage{xspace}
\usepackage{tabularx}
\usepackage{color}
\usepackage{colortbl}
\usepackage{caption}
\usepackage{subcaption}
\usepackage[colorinlistoftodos,prependcaption,textsize=tiny]{todonotes}
\usepackage{enumerate}
\usepackage[shortlabels]{enumitem}
\usepackage{amsmath, amsthm}

\usepackage{amssymb}
\usepackage{braket}
\usepackage[ruled,vlined,linesnumbered]{algorithm2e}
\usepackage{xcolor}

\definecolor{light-gray}{gray}{0.8}
\definecolor{codegreen}{rgb}{0,0.6,0}
\definecolor{codegray}{rgb}{0.5,0.5,0.5}
\definecolor{codepurple}{rgb}{0.58,0,0.82}
\definecolor{backcolour}{rgb}{0.95,0.95,0.92}

\newcommand{\tool}[1]{\textsc{#1}\xspace}

\newcommand{\bugsphp}{\tool{BugsPHP}}

\usepackage{tcolorbox}

\usepackage{multirow}
\usepackage{listings}
\usepackage{etoolbox,environ}
\newtoggle{isArxivVersion}

\definecolor{fpbackcolor}{RGB}{242,242,242}
\definecolor{diffrem}{RGB}{202, 45, 49}
\definecolor{diffincl}{RGB}{0, 135, 90}
\definecolor{codepink}{RGB}{237, 2, 140}

\lstdefinestyle{fpstyle}{
    frame=single,
    framerule=0pt,
    backgroundcolor=\color{fpbackcolor},
    resetmargins=true,
    numbers=none,
    language=Java,
    alsoletter={>,?,:,!},
    morekeywords={!,>,?,:},
    morecomment=[l][\color{diffincl}]{+},
    morecomment=[l][\color{diffrem}]{-},
    morecomment=[is][\color{diffincl}]{+*}{*+},
    morecomment=[is][\color{diffrem}]{-*}{*-},
    texcl=false
}

\usepackage{pifont}% http://ctan.org/pkg/pifont

\usepackage{amsthm}

\def\showxheight#1{
  \font\fontfoo=#1 at 10pt
  \message{The x-height of #1 at 10pt is \the\fontdimen5\fontfoo}}

\usepackage[referable]{threeparttablex}
\usepackage{soul,xcolor}
\setstcolor{blue}

\usepackage{xcolor}
\usepackage{soul}

\definecolor{lightShade}{HTML}{EFEFEF}
\definecolor{darkShade}{HTML}{D9D9D9}

\newcommand{\blueline}{\raisebox{2pt}{\tikz{\draw[-,black!10!blue,solid,line width = 0.9pt](0,0) -- (5mm,0);}}}

\newcommand{\redline}{\raisebox{2pt}{\tikz{\draw[-,red,solid,line width = 0.9pt](0,0) -- (5mm,0);}}}

\begin{document}
\title{\bugsphp: A dataset for Automated Program Repair in PHP}


% \author{\IEEEauthorblockN{Anonymous Author(s)}}

\author{K.D. Pramod}
\affiliation{%
  \institution{University of Moratuwa}
  \city{Moratuwa}
  \country{Sri Lanka}}
\email{dushan.18@cse.mrt.ac.lk}

\author{W.T.N. De Silva}
\affiliation{%
  \institution{University of Moratuwa}
  \city{Moratuwa}
  \country{Sri Lanka}}
\email{thushann.18@cse.mrt.ac.lk}

\author{W.U.K. Thabrew}
\affiliation{%
  \institution{University of Moratuwa}
  \city{Moratuwa}
  \country{Sri Lanka}}
\email{udithk.18@cse.mrt.ac.lk}

\author{Ridwan Shariffdeen}
\authornote{corresponding author}
\affiliation{%
  \institution{National University of Singapore}
  \city{Singapore}
  \country{Singapore}}
\email{ridwan@nus.edu.sg}

\author{Sandareka Wickramanayake}
\affiliation{%
  \institution{University of Moratuwa}
  \city{Moratuwa}
  \country{Sri Lanka}}
\email{sandarekaw@cse.mrt.ac.lk}

\begin{abstract}
% Automated Program Repair (APR) improves developer productivity by saving debugging and bug-fixing time. 
% %APR has been explored for C/C++ and Java programs while extended for other programming languages such as Python, JavaScript, and Haskell. Nevertheless, developing APR tools for PHP programs has not been explored much, mainly due to the nonexistence of a benchmark PHP bugs dataset. 
% %While APR has been implemented for programming languages like C/C++, Java, Python, JavaScript, and Haskell, it has not been extensively applied to PHP programs due to a lack of benchmark PHP bug datasets. 
% PHP has been the most used server-side language for over two decades and has been used in various contexts, including e-commerce, social networking, and content management. Yet, there is little to no research on bugs in PHP programs. In this work, we present a PHP bug dataset to enable research on analysis, testing, and repair for PHP programs. This paper presents \bugsphp, a benchmark dataset of PHP bugs on real-world applications. The dataset consist of training and test data-set, separately curated from GitHub and processed locally. The training data-set includes more than 600,000 bug fixing commits. The test data-set includes 513 manually validated bug-fixing commits equipped with developer-provided test cases as oracles for patch correctness assessment. 

%================Sandareka's version
Automated Program Repair (APR) improves developer productivity by saving debugging and bug-fixing time. While APR has been extensively explored for C/C++ and Java programs, there is little research on bugs in PHP programs due to the lack of a benchmark PHP bug dataset. This is surprising given that PHP has been one of the most widely used server-side languages for over two decades, being used in a variety of contexts such as e-commerce, social networking, and content management. This paper presents a benchmark dataset of PHP bugs on real-world applications called \bugsphp, which can enable research on analysis, testing, and repair for PHP programs. The dataset consists of training and test datasets, separately curated from GitHub and processed locally. The training dataset includes more than 600,000 bug-fixing commits. The test dataset contains 513 manually validated bug-fixing commits equipped with developer-provided test cases to assess patch correctness.

%\bugsphp consists 513 bugs from 15 PHP applications with a developer test suite consisting of at least one failing test case with a bug-fixing commit for correctness evaluation purposes.

% This paper presents \bugsphp, a dataset of PHP bugs on real-world applications with developer-provided test cases as oracles for patch correctness assessment. \bugsphp consists of a training dataset of 653,606 PHP bugs and a test dataset of 513 bugs. The \bugsphp test dataset includes a developer test suite consisting of at least one failing test case with a bug-fixing commit for correctness evaluation purposes. 

% Further, we evaluate the efficacy of Large Language Models:\textit{TextDavinci} and \textit{ChatGPT} in fixing PHP bugs using \bugsphp dataset. Our results demonstrate that TextDavinci outperforms ChatGPT in generating more plausible patches. On average, TextDavinci produces patches of higher quality, equivalent to those made by developers. Our results suggest that further study of PHP bugs is required to understand the unique complexities of PHP language that are uncommon with other programming languages. Our contribution is an attempt to foster further research on PHP applications.

\end{abstract}

\keywords{Automated Program Repair, PHP Application Errors}

\maketitle

\section{Introduction}
Manually fixing software bugs is a tedious and time-consuming task. Software engineers must invest significant time and effort in finding and fixing bugs in software programs. For example, O'Dell et al. in \cite{O'Dell2017} have mentioned that developers spend 50\% of programming time finding and fixing bugs. As a solution, Automated Program Repair (APR)~\cite{apr_cacm19} has been explored to reduce the bug-fixing effort and increase developer productivity. Given the buggy program, APR tools automatically generate potential patches for software bugs. APR tools analyze the buggy program to determine the root cause and create a patch that fixes the bug while preserving the original program functionality \cite{automatic_software_repair_a_survey}. Most of the existing APR tools, particularly the learning-based ones, have been implemented for bug repair in C, Java, Python, and JavaScript languages owing to the availability of large datasets for those languages. Surprisingly, despite being the most popular server-side scripting language, there is no benchmark bug dataset for PHP. %and thus developing APR tools for programs in PHP is less explored. 

%For example, datasets such as Defects4J \cite{just2014defects4j}, Bugs.jar \cite{bugsjar}, Bears \cite{madeiral2019bears} contain bugs for Java programs whereas QuixBugs \cite{lin2017quixbugs} consists of bugs in both Java and Python languages. ManyBugs and IntroClass \cite{le2015manybugs} are datasets created for C language. The datasets for JavaScript include BugsJS \cite{gyimesi2019bugsjs} and FixJs \cite{csuvik2022fixjs}. 
% Surprisingly, despite being the most popular server-side scripting language, there is no benchmark dataset for PHP. The unavailability of a benchmark dataset for PHP bugs hinders the development of APR tools for PHP program bugs. 

PHP is an open-source and general-purpose language that powers many large-scale web applications like Facebook, Wikipedia, and WordPress. Owing to the easiness of programming, PHP is still among the top choices for web development, and PHP is used by 77.5\% of websites as the server-side scripting language \cite{W3Techs}. Weak typing, poor performance, and lack of debugging tools cause errors in PHP web applications. Despite its popularity, the research community has paid little attention to developing techniques to improve PHP programs. A well-organized labeled PHP bugs dataset must be introduced to facilitate further studies of the software evolution, maintenance, and repair of PHP applications.

% In this work, we introduce a PHP bug dataset called \bugsphp that can be used to develop and evaluate APR models for PHP programs and conduct an empirical study to assess LLMs and existing APR tools on repairing PHP bugs using the introduced \bugsphp dataset. We also present a preliminary study of using two LLMs developed by OpenAI, \textit{text-davinci-003} (i.e. TextDavinci) and \textit{gpt-3.5-turbo} (i.e. ChatGPT) on \bugsphp. Our empirical study demonstrates that TextDavinci outperforms ChatGPT in repairing PHP bugs, despite ChatGPT being the more advanced LLM. For instance, TextDavinci correctly fixes 41 bugs out of 513 in our dataset, while ChatGPT only fixes 19 bugs. 

In this work, we introduce \bugsphp, a data-set of PHP programs that can be used to train deep learning models and evaluate any APR technique including traditional techniques that are not learning models. \bugsphp contains 653606 bug fixing commits and a separate set of 513 bug fixing commits equipped with at least one failing test case, by crawling 5000 GitHub repositories with PHP code. We collect commits from 01 Jan 2020 to Mar 2023. 
%Following related work, we collect commits with the number of line changes less than 50 and the number of file changes less than 3. 
For the test data-set construction, we carry out a manual test case validation by running developer provided test cases on the fixed version of the program. We select the commits with at least one failing test case from its parent commit, resulting in 513 bugs from 15 PHP applications that constitute the buggy version, fixed version, and related test cases. Excluding the project repositories used for the test data-set, the remaining projects are used to construct the training data-set. 
This paper makes three main contributions:

%In this work, we introduce a PHP bug dataset called \bugsphp that can be used to develop and evaluate APR techniques for PHP programs. \bugsphp consists of a training dataset containing 653,606 PHP bugs from 4,483 PHP applications and a test dataset containing 513 bugs from 15 PHP applications. In curating the test dataset, we first select the most popular 75 GitHub Repositories based on the stargazers count and remove those that do not have developer test-suite. Next, we collect commits from 01 Jan 2020 to Mar 2023 based on the commit messages from the filtered set of repositories. Following the existing work, we collect commits with the number of line changes less than 50 and the number of file changes less than 3. Then, we carry out a manual test case validation by running all the test cases on the fixed version of the program. Finally, we select the commits with at least one failing test case from its parent commit, resulting in 513 bugs from 15 PHP applications that constitute the buggy version, fixed version, and related test cases. We follow a similar subject selection and bug collection approach to curate the training dataset. This paper makes three main contributions

\begin{itemize}
    \item A PHP bug dataset called \bugsphp consisting of 513 PHP bugs for testing and 653606 for training deep learning models from popular open-source projects. 
    \item Analysis of the types of errors in PHP applications captured in \bugsphp
    \item Preliminary results of the effectiveness of existing APR Models to fix errors in PHP applications
\end{itemize}

% In the later sections, this paper discussed our methodology to construct the dataset of \bugsphp, a dataset representation, and the result we obtained after evaluating with LLMs.

\section{Motivation}
PHP is the most popular server-side language used by major web applications, including Facebook, Wikipedia, Tumblr, Slack, MailChimp, Etsy, and WordPress. Facebook alone boasts 2.9 billion users globally, while Wikipedia has received around 5 billion visits in the first 5 months of 2022. Slack is expected to have 32.3 million monthly users by 2023, and WordPress has been used to create 810 million websites, accounting for 43\% of all websites. 

However, like any other software, PHP programs can have bugs ranging from minor issues to major security flaws. These bugs can lead to severe problems such as data loss or unauthorized access by attackers who exploit vulnerabilities in web applications. For instance, the 2017 Equifax data breach \cite{equifax_data_breach} cost the company \$700 million in expenses and lost revenue due to a flaw in the company's web application and affected nearly 150 million clients, about 56\% of Americans. 87 million records of US resident profiles were obtained through Facebook as the most significant data breaches on online social networks as of 2020, and the data was used to create software that could predict and influence electors \cite{online_social_networks_security_and_privacy}.

% Since billions of internet users rely on web applications developed in PHP, any bugs or issues in PHP web applications can significantly impact many users. Hence, it is vital to investigate bugs that may occur in PHP applications. By analyzing these bugs, we can improve the stability and security of web applications and increase the effectiveness and efficiency of software development. In addition, studying PHP vulnerabilities can help identify critical issues, develop effective mitigation plans, and prevent costly data breaches and security incidents that damage a company's reputation.

Nevertheless, manually fixing program bugs can be costly and time-consuming, causing delays in releasing new features or products. To save resources and time, developers can use efficient tools and methods to identify and resolve these issues, enabling them to focus on creating high-quality software. To better understand the nature of bugs in PHP applications, there is a need for a standard dataset to study the different types of bugs and their required fixes. Additionally, a comprehensive dataset can facilitate the analysis of patterns and traits of PHP bugs over time, investigating their effects on web application security and developing innovative software testing and verification methods.

This work aims to fill this gap by curating a list of PHP bugs from the most popular open-source PHP applications. Researchers can use our work to evaluate program repair tools for PHP applications. Furthermore, this dataset can be applied to identify insecure code, detect code smells and create new methods and tools for debugging PHP bugs. Overall, this research aims to benefit the community by offering a valuable dataset of PHP bugs. % identification and prevention, enhancing the reliability and security of PHP applications.

\section{Related Work}
The existing work in analyzing program bugs has curated datasets focused on Java, C, Python, and JavaScript bugs. For example, Defects4J\cite{just2014defects4j} is a framework and database that offers genuine Java program bugs for reproducible research in software testing. It has 357 bugs from five open-source programs, each with a complete test suite that exposes the bug. It's expandable, and new bugs can be easily added to the database once a program is set up. Bugs.jar \cite{bugsjar} is an extensive data collection useful for researching automated Java program debugging, testing, and patching. It contains 1,158 bugs and patches from 8 open-source Java projects representing eight important application categories. BEARS~\cite{madeiral2019bears} is a project that creates a flexible benchmark for automatic repair studies in Java. It collects and stores bugs by identifying potential pairs of faulty and fixed program versions from open-source projects on GitHub. BEARS is publicly accessible and includes 251 reproducible bugs from 72 projects that use the Travis CI and Maven build environment.

The ManyBugs and IntroClass datasets \cite{le2015manybugs} comprise 1,183 defects found in 15 C/C++ programs. %These datasets help evaluate automatic repair algorithms and come with reproducibility and benchmark quality guarantees. 
ManyBugs contains bugs from well-known open-source projects, while IntroClass contains bugs from programming assignments done by a small group of students. The datasets BugsJS \cite{gyimesi2019bugsjs} and FixJS \cite{csuvik2022fixjs} are collections of JavaScript bugs. BugsJS features 453 actual bugs that have been verified manually. These bugs are taken from 10 widely used server-side JavaScript programs, which collectively have 444,000 lines of code. FixJS, on the other hand, gathers bugs from GitHub and provides information on the faulty and fixed versions of the same program. This dataset comprises over 300,000 samples and includes details on the commit, before/after states, and three source code representations. BugsInPy~\cite{bugsinpy} is a dataset that collects real world bugs from Python programs, where each is accompanied by
a failing test case that passes once the bug is fixed. Vul4J~\cite{vul4j} is a more recent work collecting vulnerability fixing commits for Java programs. 
%Recently, Julian Aron Prenner et al. in \cite{prenner_robbes_2023}  present RunBugRun, a fully executable dataset of 450,000 small buggy/fixed program pairs originally submitted to programming competition websites written in eight different programming languages, namely C, C++, Java, Python, JavaScript, Ruby, Go and PHP. All the bugs in the RunBugRun dataset come with test cases. Further, the authors evaluate the dataset with two baselines and provide some statistical information. 

% In contrast, \bugsphp is generic and contains information on PHP bugs. Unlike previous datasets, we provide support for both learning-based APR techniques (by providing training dataset) and traditional APR techniques (by providing test cases for patch validation). 

In contrast, \bugsphp is generic and contains information on PHP bugs. Unlike previous datasets, we provide support for learning-based and traditional APR techniques by providing a training dataset and test cases for patch validation. 

% The existing learning-based APR methods that involve training machine learning models with a large corpus of code and natural language descriptions. These methods can generate semantically meaningful patches. However, most of the existing APR tools focus on C, Java, JavaScript, and Python languages. For example, SequenceR \cite{chen2019sequencer}, RewardRepair \cite{ye2022neural}, and CURE \cite{jiang2021cure} are APR tools developed for Java, whereas BIFI \cite{yasunaga2021break}, DrRepair \cite{yasunaga2020graph} and RSRepair \cite{qi2014strength} are developed for C programs. See Table~\ref{tab:list-of-apr-tools} shows a list of APR tools categorization based on the focused programming language.

% However, no learning-based APR tools are available for PHP bugs, and the existing tools cannot repair logical bugs. Samimi et al. in \cite{samimi2012automated} introduce two PHP and HTML error-repairing tools for PHP applications. The first tool, PHPQuickFix, analyzes individual prints to fix basic bugs. The second tool, PHPRepair, uses a dynamic approach to handle more complex repairs. Fix Me Up \cite{son2013fix} is a tool designed to repair access-control bugs in PHP web applications. It works through static analysis and transformation, using an Access Control Template to identify flawed access control logic in access check statements. Once identified, Fix Me Up inserts missing statements to fix the bugs.
\begin{figure*}[!t]
    \centering
    \includegraphics[width=0.85\linewidth]{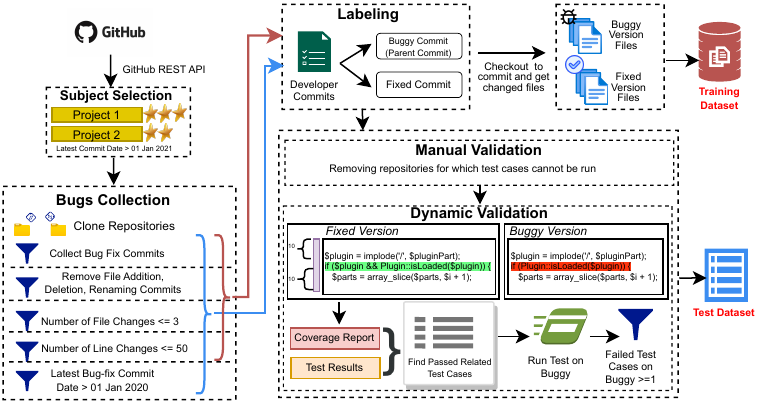}
    \caption{The overview of the process of curating \bugsphp dataset.}
    \textbf{The blue line \blueline~ and red line \redline~ represent the filtering criteria for the test data-set and training data-set, respectively. }
    \label{fig:benchmark_flow}

\end{figure*}

\section{\bugsphp Dataset}
In this section, we present the methodology we followed to curate a dataset of PHP bugs and provide a statistical overview. Our new dataset, \bugsphp, includes a training dataset of 653,606 PHP bug-fixing commits and a test dataset of 513 bugs from the most popular open-source applications collected from GitHub between 01 Jan 2021 and March 2023. %This dataset can be valuable for evaluating software testing, program analysis techniques, and program repair methods.

\subsection{Methodology}

The process of curating the \bugsphp dataset is outlined in Figure~\ref{fig:benchmark_flow}. Based on their stargazers count, we use the GitHub REST API to retrieve the top 5000 PHP repositories with at least one commit after 01 Jan 2021 (indicating recent development activities). We avoid repositories that are not maintained by filtering those which do not have recent development activities. 
Bugs are then collected from these repositories and filtered based on the number of file and line changes they have. After filtering, the top 75 repositories' bugs are used for the testing dataset, while the remaining repositories are used for the training dataset. For the bugs in the training dataset, we extract the buggy and fixed versions files for the \bugsphp training dataset. Meanwhile, we conduct a dynamic validation for the bugs in the testing dataset to obtain each bug's corresponding buggy version, fixed version, and test cases. Below, we elaborate on each step of curating our dataset.

\textbf{Repository Selection:} To collect commits, we retrieve PHP repositories using the GitHub REST API \footnote{https://docs.github.com/en/rest?apiVersion=2022-11-28}. Then, we sort them by popularity, measured using the stargazers count (e.g., the number of stars) and the latest commit date. We select only repositories with at least one commit date after 01 Jan 2021 to ensure that we collect bugs from recently updated repositories. Finally, we collect commits from the top 5,000 repositories.

\textbf{Bug Collection:} To collect bugs, we search for PHP projects on GitHub and use the git version control system's history to locate issues and the corresponding solutions developers provided. We follow previous studies (e.g., ~\cite{tufano2019empirical}) and look for commits with messages containing words like ``fix", ``solve", ``bug", ``issue", ``problem," or ``error". Finally, we apply the following criteria similar to those in~\cite{gyimesi2019bugsjs} to filter out unusable data. a) developer fix is applied only to PHP files b) developer patch should not contain file addition, deletion, or renaming, as such modifications to the source files are usually refactoring effects and can obfuscate the relevant fix c) number of file changes should be three or fewer and d) number of line changes should be equal to or less than 50. We then sort through repositories containing the bugs that meet the criteria and have bug fixes committed after 01 Jan 2020. Then, we choose the top 75 most popular repositories from this list to create the \textit{\bugsphp testing dataset}. The remaining repositories are used to select bugs for the \textit{\bugsphp training dataset}. 

% \textbf{% Please add the following required packages to your document preamble:
% \usepackage{multirow}
% \usepackage{graphicx}
\begin{table*}[]
\centering
\caption{Overview of BugsPHP test dataset of PHP applications. BugsPHP tests were taken from the fix commit of the subject, which corresponds to a passing test-case with the fix-commit. The test coverage columns summarize the coverage information typically having a high coverage in terms of lines and functions. }

\label{table:data-set}
\resizebox{\textwidth}{!}{%
\begin{tabular}{@{\extracolsep{4pt}}lllllrrrlrr}

\hline

\multicolumn{2}{c}{\textbf{Subject}} & \multicolumn{3}{c}{\textbf{Repository Information}} & \multicolumn{3}{c}{\textbf{Test Coverage}} & \multicolumn{1}{c}{\multirow{2}{*}{\textbf{LOC}}} & \multicolumn{1}{c}{\multirow{2}{*}{\textbf{\# Bugs}}} & \multicolumn{1}{c}{\multirow{2}{*}{\textbf{\# Tests}}} \\ \cline{3-5}  \cline{6-8}

\multicolumn{1}{c}{\textbf{Name}} & \multicolumn{1}{c}{\textbf{Description}} & \multicolumn{1}{c}{\textbf{Stars}} & \multicolumn{1}{c}{\textbf{Forks}} & \multicolumn{1}{c}{\textbf{Commits}} & \multicolumn{1}{c}{\textbf{Lines}} & \multicolumn{1}{c}{\textbf{Functions}} & \multicolumn{1}{c}{\textbf{Classes}} & \multicolumn{1}{c}{} & \multicolumn{1}{c}{} & \multicolumn{1}{c}{} \\

\hline
\hline

cakephp & web application framework & 8.6k & 35k & 44.6k & 84.69\% & 73.59\% & 36.92\% & 358k & 33 & 8322 \\
carbon & API extension for DateTime & 16.1k & 12k & 3.4k & 99.98\% & 99.75\% & 98.25\% & 316k & 11 & 5898 \\
composer & dependency manager & 27.6k & 65k & 12.0k & 63.34\% & 49.35\% & 14.53\% & 107k & 18 & 2240 \\
dbal & database abstraction layer & 9k & 12k & 10.9k & 65.52\% & 55.72\% & 29.28\% & 81k & 9 & 4122 \\
easywechat & WeChat API for PHP applications & 10k & 24k & 2.1k & 56.02\% & 50.00\% & 20.25\% & 14k & 9 & 178 \\
framework & web application framework & 29.3k & 99k & 35.4k & 74.91\% & 69.09\% & 32.28\% & 355k & 94 & 8425 \\
google-api-php-client & Google API for PHP applications & 8.5k & 35k & 1.8k & 66.60\% & 55.56\% & 11.76\% & 11k & 3 & 228 \\
laravel-permission & permission manager & 11.2k & 17k & 1.3k & 93.72\% & 80.29\% & 53.85\% & 8k & 6 & 432 \\
magento2 & e-commerce platform & 10.6k & 92k & 135.3k & 46.15\% & 37.66\% & 29.52\% & 2854k & 23 & 16224 \\
monolog & library for logging & 20.3k & 19k & 2.6k & 63.98\% & 52.97\% & 23.85\% & 27k & 7 & 1139 \\
orm & object relation mapper & 9.6k & 25k & 13.1k & 84.27\% & 68.01\% & 45.09\% & 187k & 15 & 3706 \\
PHP-CS-Fixer & linter for coding standards & 11.9k & 15k & 8.8k & 93.97\% & 87.45\% & 64.85\% & 225k & 82 & 31774 \\
PHP-Parser & static analyzer for PHP & 16.1k & 0.9k & 1.4k & 92.69\% & 91.29\% & 85.90\% & 30k & 3 & 1691 \\
PhpSpreadsheet & library for spereadsheet files & 12.1k & 30k & 4.0k & 80.01\% & 77.21\% & 52.25\% & 251k & 12 & 14071 \\
symfony & web application framework & 28.2k & 90k & 63.9k & 80.66\% & 65.67\% & 36.09\% & 1579k & 188 & 41083 \\

\hline
\hline
\multicolumn{2}{c}{Total / Average} & 229.1k & 570.9k & 340.6k & 76.43\% & 67.57\% & 42.31\% & 6.4M & 513 & 139k \\
\hline
\end{tabular}%
}
\end{table*}

\textbf{Dataset Labeling:} We obtain the developer commits from the set of repositories selected and designate the commit before the bug-fixing commit as the \textit{buggy version} of the program and the current version as the \textit{fixed version} of the program. Based on the labelled developer commits, we extract the corresponding buggy version and fixed version files from the repositories designated to obtain the training commits to creating the \textit{\bugsphp training dataset}. The labelled developer commits obtained from the test dataset repositories are fed to the validation process. In creating the \textit{\bugsphp testing dataset}, we perform manual and dynamic validations on the bugs selected for the testing dataset.

\textbf{Manual Validation:} In this step, we examine the latest bug to validate a repository and see if its fixed version has accompanying test cases that can verify the fix. We use the latest stable PHP version, 8.1, to validate fixed versions with these test cases. We selected PHP version 8.1 since it is the latest stable PHP version at the time of writing. If the latest bug passes this step, we move that repository to the dynamic validation. If not, we repeat the process with the following bug with the next lower PHP version until we find a bug that passes. This validation is done for each repository until we find at least one bug that successfully passes the validation. Thus, we remove repositories for which we do not find at least one bug that can successfully run the validation. 

\textbf{Dynamic Validation:} We run all the test cases on the fixed version of the program and generate the test coverage report. Then, using the test coverage report, we identify the relevant test cases that reach the fixed version's modified location(s). Next, we select the appropriate passing test cases covering the modified fix locations within a neighborhood of 10 lines. Next, we run selected relevant test cases on the buggy version of the program to identify at least one test case that fails for that buggy version. Finally, we choose the bugs with at least one failing test case. These validation steps result in a testing dataset of 513 bugs, each consisting of the buggy version, the fixed version, and the corresponding test cases. The details of the \textit{\bugsphp testing dataset} are shown in Table~\ref{table:data-set}.

\textbf{Training Data}: We also curated a training dataset following similar criteria for subject selection and bug collection in bug dataset construction. In Subject selection, we used GitHub REST API\footnote{https://docs.github.com/en/rest?apiVersion=2022-11-28} to retrieve the GitHub repositories. Following steps similar to those previously taken, we selected the first 5k repositories and removed the repositories included in our bug dataset to curate a training dataset. In the bugs collection step, we collected the bug fix commits by checking \texttt{"fix"}, \texttt{"solve"}, \texttt{"bug"}, \texttt{"issue"}, \texttt{"problem"} or \texttt{"error"} keywords included in the commit message. Then, we filtered the bug fixes with the number of file changes less than or equal to 3, and the number of line changes less than or equal to 50. Then, we extracted the changed files and labeled them buggy and fixed. We collected 653,606 bugs from 4483 PHP applications. Finally, we included a metafile that contains the commit ID, repository, changed line numbers, etc, for each data point.

\subsection{Overview of the \bugsphp Dataset}
We analyzed the size of the fix commits in terms of the number of line and file changes to understand the complexities of the fix commits. Most commits modify only 1 file, with 76.3\% of the training dataset (498940 fix commits) and 92.2\% of the testing dataset (473 bugs) representing single file changes. The percentage of fix commits with 2 and 3 file changes in the training dataset are 14.2\% and 9.5\%, respectively. In the test dataset, 5.8\% of the dataset (30 bugs) have two file changes, whereas 2\% of the dataset (10 bugs) have three file changes. A similar pattern exists for the number of lines of the commits. In the training dataset, 67.1\% of commits have 1-10 line changes; in the testing dataset, 68\% of bugs fall within this range. Also, the training dataset contains 101776 fix commits (15.6\%)  with 11-20 line changes, 50195 fix commits (7.7\%) with 21-30 line changes, 31081 fix commits (4.8\%) with 31-40 line changes, and 30760 fix commits (4.8\%) 41-50 line changes.

In addition to the size of the patches, we also analyzed the types of bugs in our test dataset. We identified 462 bugs (90\%) as functional errors, while the remaining bugs consist of 16 type errors, 15 security vulnerabilities, 13 compatibility issues, 5 usability issues and 2 performance bugs.

\section{Preliminary Results with APR}
% We trained two learning based APR models, CURE~\cite{jiang2021cure} and RewardRepair~\cite{ye2022neural} with our training data and evaluated using 513 bugs in our test data. For each model we generated 100 candidate patches per bug. Table~\ref{tab:apr-tools-patch-results} shows the patch result of each model. RewardRepair generates a patch for 103 bugs which pass the failing test cases, but CURE generated only for 48 bugs. Out of these bugs RewardRepair fixed 43 bugs and CURE fixed 11 bugs, which passes all test cases. RewardRepair could generate 3-4 line  patches, which CURE only generated single line patches. RewardRepair fixed 29 unique while CURE fixed 4 unique bugs.  Both APR models commonly fixed 7 bugs. However, none of the APR models could fix 473 bugs in our test dataset. 

% We show that our data can be directly used by existing APR models and further research is needed to improve the capabilities of existing APR for PHP bugs. 

With our training data, we train two learning-based APR models, CURE~\cite{jiang2021cure} and RewardRepair~\cite{ye2022neural}, and evaluate using 513 bugs in our test data. For each model, we generate 100 candidate patches per bug. Table~\ref{tab:apr-tools-patch-results} shows the patch result of each model. Columns $N_C$, $N_V$ and $N_P$  depicts \# bugs a candidate patch was generated, failing test cases are fixed, and that pass all test cases, respectively. Similarly columns $N_E$ and $N_I$ depicts \# bugs with a semantically equivalent patch and a identical patch, respectively.

\begin{table}[H]
\centering
\small
\caption{Efficacy of APR models in \bugsphp}
\begin{tabular}{|c|c|c|c|c|c|}
\hline
\textbf{Model} & \textbf{$N_C$} &
\textbf{$N_V$}& \textbf{$N_P$} &
  \textbf{$N_E$} & \textbf{$N_I$} \\ \hline
CURE &
  443 &
  48 &
  11 &
  1 &
  0 \\ \hline
RewardRepair &
  513 &
  103 &
  43 &
  13 &
  6 \\ \hline
\end{tabular}
\label{tab:apr-tools-patch-results}
\end{table}

RewardRepair generates a patch for 103 bugs that pass the failing test cases, but CURE generates patches only for 48 bugs. RewardRepair fixes 43 bugs out of these bugs, and CURE fixes 11 bugs, which passed all test cases. RewardRepair can generate 3-4 line patches, while CURE only generates single line patches. RewardRepair fixes 29 unique while CURE fixes four unique bugs. Both APR models commonly fixed seven bugs. However, none of the APR models could fix 473 bugs in our test dataset. We show that existing APR models can be evaluated using our new data set, and further research is needed to improve the capabilities of existing APR for PHP bugs.

\section{Conclusion}
PHP has been the most popular server-side language for over two decades, yet there is no dataset to study bugs appearing in PHP applications. Hence, we have curated a dataset of bug-fixing commits from the most popular open-source PHP applications to train learning-based repair tools and evaluate program repair techniques. \bugsphp  contains a training dataset of 653,606 bug-fixing commits and a test dataset of 513 bugs, which maintains developer-written tests using the PHPUnit testing framework. 

% We performed an empirical study in fixing logical errors of PHP applications using the test dataset of \bugsphp. For this purpose, we leveraged two Large Language Models: \textit{TextDavinci}, \textit{ChatGPT}, and two state-of-the-art learning-based program repair tools: \textit{CURE} and \textit{RewardRepair}. Our study indicates program repair tools have competitive performance in finding a correct patch compared to pre-trained Large Language models, which have been trained on vast amounts of data. Furthermore, the number of bugs fixed by repair techniques and large language models is less than 10\% of the total bugs in our dataset, indicating that further research is needed to improve the state-of-the-art program repair techniques for PHP errors.

\begin{tcolorbox}[boxrule=1pt,left=1pt,right=1pt,top=1pt,bottom=1pt]
\textbf{Data Set:}
Our dataset can be accessed via GitHub from the following repository: https://github.com/bugsphp/bugsPHP.git
\end{tcolorbox}

\section*{Acknowledgement}
This work was partially supported by a Singapore Ministry of Education (MoE) Tier3 grant “Automated Program Repair”, MOE-MOET32021-0001.

\bibliographystyle{ACM-Reference-Format}
\bibliography{references}

\end{document}